\documentclass[12pt,fleqn]{article}
\usepackage[dvips]{color}
\usepackage[dvips]{graphicx}

\begin{document}
\title{Superluminal Tunneling Devices}

\author{ G\"unter  Nimtz\\
II. Physikalisches Institut, Universit\"at K\"oln}

\date{November 2001}
\maketitle

\begin{abstract}
Photonic tunneling permits superluminal signal 
transmission. The principle of causality is not violated but the time duration between cause and effect 
can be  shortened compared with an interaction exchange with velocity of light.   
This outstanding property can be applied to speed-up photonic signal modulation and 
transmission as well as to improve micro-electronic devices. 
Superluminal photonic signal transmission have been presented at microwave and infrared frequencies already.  
Presumably superluminal photonic and electronic devices can become reality having in mind the 
experimental evidence of the universal tunneling time of photons and of electrons.
\end{abstract}

\section{Introduction}
In 1992 Enders and the author  have demonstrated that photonic 
tunneling takes place with superluminal signal velocity~\cite{Enders}. The experiments were carried out 
with microwaves. 
At that time any application of superluminal tunneling was not expected in spite of the popular  
semiconductor tunneling diode. 
A decade later I am going to present some potential applications of superluminal tunneling in photonics and electronics.

The special features of evanescenuniversal tunneling tit modes and wave mechanical tunneling are presented in this chapter. In the following 
chapters the essential  properties of a signal as well as the universal tunneling time are introduced. The last chapters are devoted 
to applications of the tunneling process as well as to a 
summary of the strange tunneling properties.

\begin{figure}
    \includegraphics[width=\linewidth]{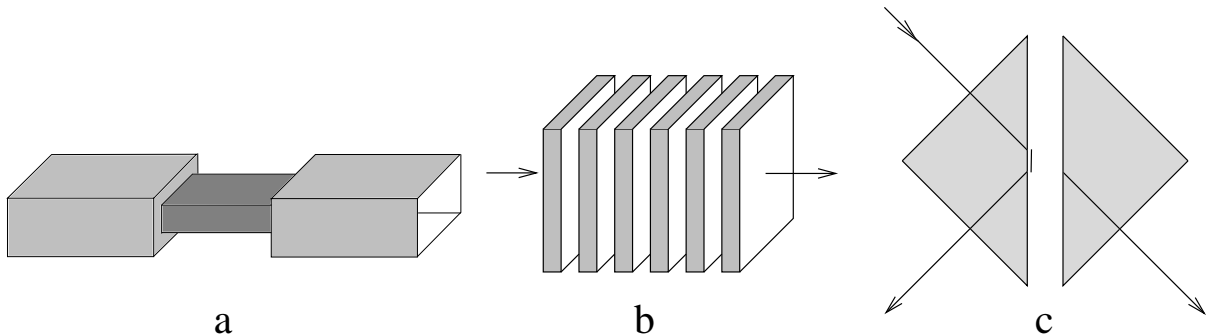}
\caption{Sketch of  three prominent photonic barriers. a) illustrates an undersized wave guide 
(the central part of the wave guide has a cross-section being smaller than half the wavelength 
in both directions perpendicular to propagation) , b) a photonic lattice (periodic dielectric hetero 
 structure), and c) the frustrated total internal reflection of a double prism, where total reflection 
takes place at the boundary from a denser to a rarer dielectric medium.}
\label{Barrier}
\end{figure}

Tunneling is the wave mechanical presentation of evanescent modes~\cite{Sommerfeld,Feynman}. Evanescent modes are
dominantly found 
in undersized waveguides, in the forbidden frequency bands of periodic dielectric
heterostructures, and with double prisms in the case of frustrated total reflection ~\cite{Nimtz1,Chiao}. 
These prominent examples of photonic tunneling barriers are sketched in
Fig. \ref{Barrier}.
The dielectric lattice is equivalent 
to the electronic lattice of semiconductors with forbidden energy gaps. 

Evanescent modes and tunneling wave functions are characterized by a purely ${\it  
imaginary}$ wave number. For instance the wave equation yields 
for the electric field $E(z)$
\begin{eqnarray}
E(z) =  E_0 \, e^{i(\omega t - k x)} \,\,\,\Rightarrow \,\,\,\
E(z) = E_0
\, e^{i \omega t - \kappa x} ,
\end{eqnarray}

where $\omega$ is the angular frequency, $t$ the time, $x$ the distance,
$k$ the wave number, and $\kappa$ the {\it imaginary} wave number
of the evanescent mode. 

\begin{figure}[htb]
\begin{center}
\includegraphics[width=1\textwidth]{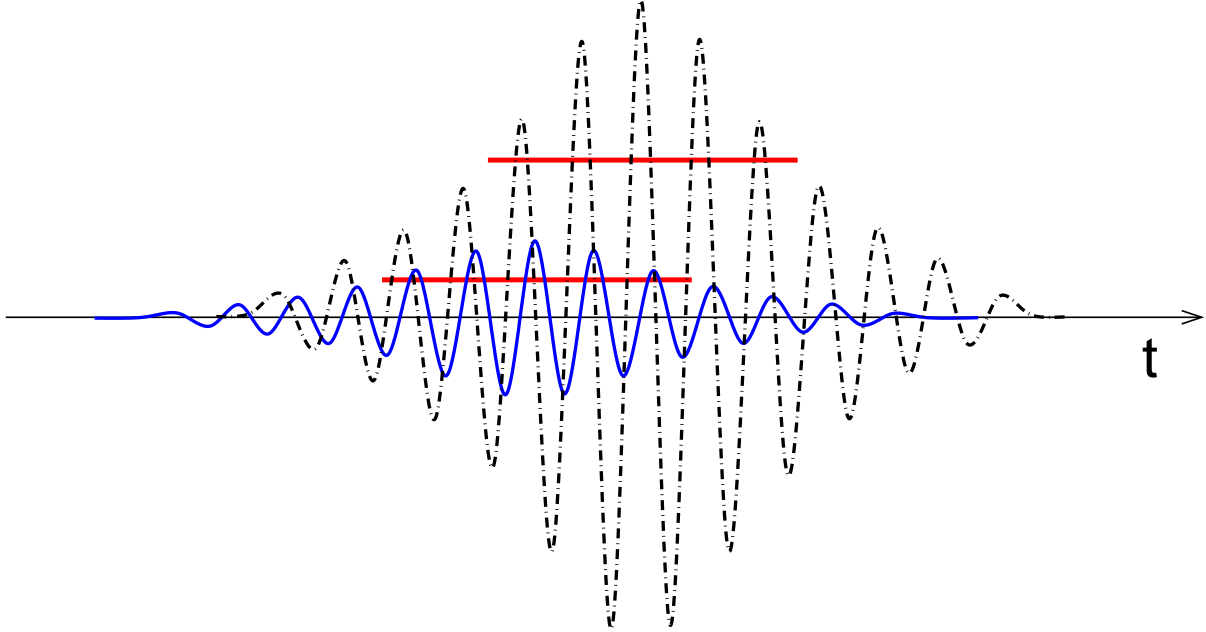}
\end{center} 
\caption{Sketch of  two wave packets (i.e. pulses), amplitude vs time. The larger packet travelled slower than 
the attenuated one. The horizontal bars indicate the half width of the packets which do not depend on the packet's magnitude. 
The figure illustrates the gradually beginning 
of the packets. The forward tail of the smooth 
envelope may be described by the relation $[1 - exp(-t/\tau)] [sin(\omega t)]$ for instance, where $\tau$ is a time constant. 
\label{Paket}}
\end{figure}

In the three introduced  examples the modes are characterized by an exponential attenuation of transmission 
due to reflection by photonic barriers and by a lack of a phase shift inside the barrier.  
The latter means a zero time barrier traversal according to
the phase time approach

\begin{eqnarray}
 \tau = d\phi/d\omega,
\end{eqnarray}

 where $\tau, \phi, \omega$
are the phase-time, the phase, and the angular frequency, respectively. The observed very short tunneling time 
is caused at the barrier front boundary. 
In this report the  tunneling time is defined as the time a group or a signal 
spent traversing a barrier. The time is measured outside the barrier with detectors positioned at the front and the back of the barrier. 

This time corresponds to the phase-time or group delay, 
see Refs.~\cite{Papoulis,IEEE,NTIA} for example.

In 
Fig.\ref{Paket} a pulse (i.e. a wavepacket) is sketched which represents a digital signal. 
The front of the envelope is very smooth corresponding to a narrow frequency band width.   
The frequency band is choosen with respect to the barrier in question 
in such a way that the pulse contains essentially evanescent frequency components.  
Such an evanescent pulse travels in zero time through 
opaque barriers, which in turn results
in an infinite velocity in the phase-time approach neglecting the phase
shift at the barrier front~\cite{Enders,Hartman}. 

In the review on {\it The quantum mechanical tunnelling time problem - revisited} by 
Collins et al. \cite{Collins},  the following statement has been made: {\it  the phase-time-result originally obtained by Wigner and 
by Hartman is the best expression to use for a wide parameter range of barriers, energies and wavepackets.} The 
experimental results of photonic tunneling have confirmed this statement ~\cite{Nimtz1}. 
 
Einstein causality prohibits superluminal signal velocity in vacuum and in media with a real refractive index. This rule 
does not hold for media characterized by an imaginary refractive index, where the phase shift is zero as in the case of 
evanescent modes.  In order to avoid signal reshaping due to the dispersion of the special media 
the signal has to be frequency band limited. 
The problem of 
limited frequency band and the 
time limited duration of signals has been discussed and explained by quantum 
mechanical arguments~\cite{Nimtz4,Nimtz5}. Actually, 
signals of limited frequency band and limited time duration have 
been transmited in the multiplex telephony for more than a hundred years. 
The theory for these technical signals and more general for all physical signals, presents the sampling theorem, 
which  has been introduced by 
Shannon around 50 years ago~\cite{Shannon}. More details about signal properties are presented in Chapter Signals. 

Evanescent modes are solutions of the classical 
Maxwell equations, however, they 
display some  nonclassical properties as for instance:

\begin{enumerate}
\item The evanescent modes seem to be represented by nonlocal fields as was predicted and later 
shown by transmission and partial reflection experiments~\cite{Nimtz1,Hartman,Vetter}. 
Tunneling and reflection times are equal and independent of barrier length.
 
\item Evanescent modes have a negative energy, thus they cannot be measured~\cite{Bryngdahl,Nimtz3}. 

\item Evanescent modes can be described by virtual photons \cite{Carniglia}.

\item Evanescent modes are not Lorentz invariant as $(v_\varphi/c)^2 ~\rightarrow~ \infty$ holds, where $v_\varphi$ = x/$\tau$ 
is the phase time velocity and c the velocity of light in vacuum. x represents the barrier length. 

\end{enumerate}

Obviously, evanescent modes are not fully describable by the Maxwell equations and 
quantum mechanics have to be taken into consideration.  This is similar to the photoelectric effect which 
Einstein explained  by quantum mechanics, i.e. by photons. In general 
electric fields are only detectable by a quantized energy exchange.

\begin{figure}[hbt]
\begin{center}
\includegraphics[width=0.4\linewidth]{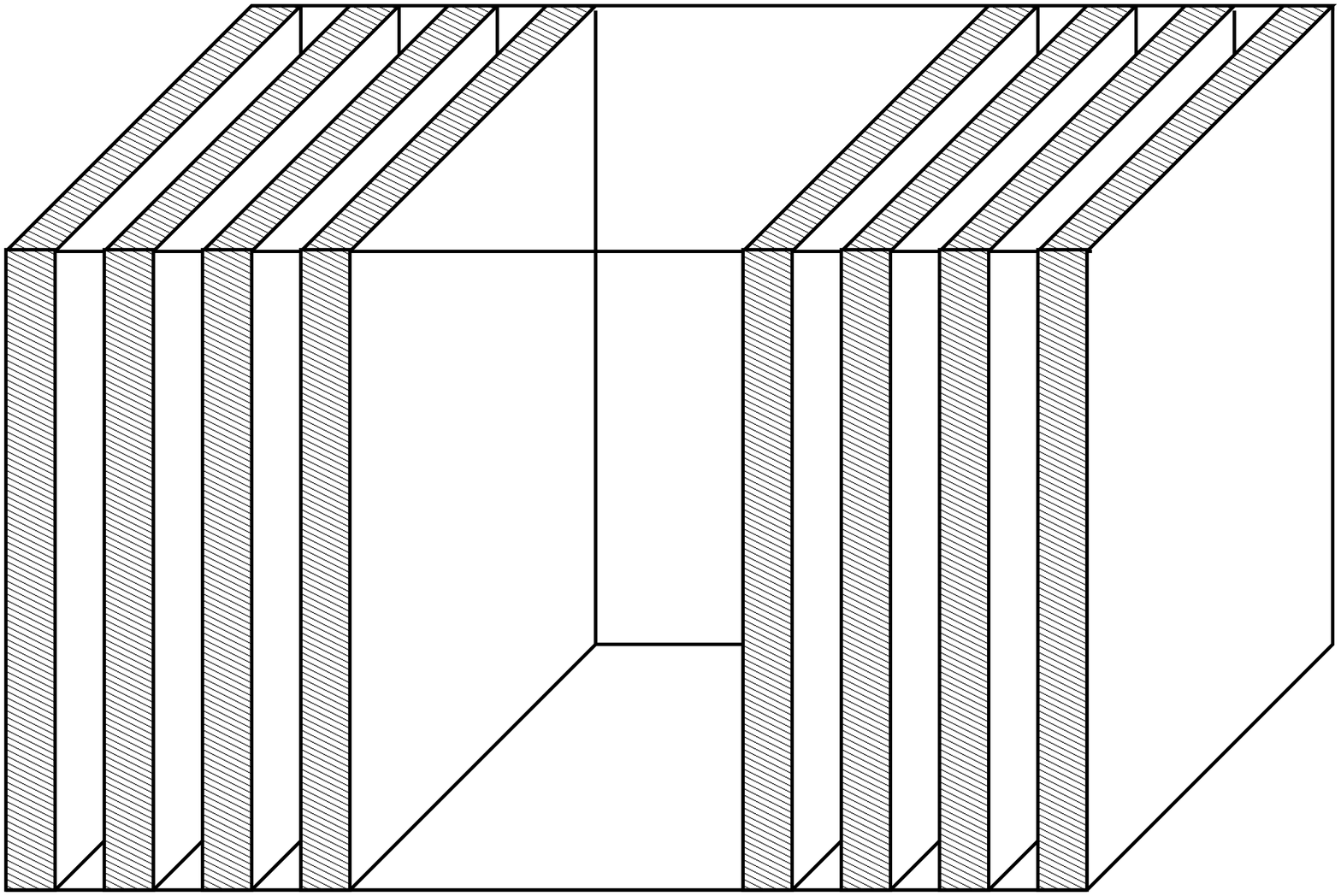}
\end{center}
\caption[]{
An example of a resonant electromagnetic tunneling structure with evanescent mode solutions (forbidden frequency bands) at microwave frequencies. 
Two periodic quarter wavelength hetero--structures
of perspex and air which are separated by a distance of 18.9~cm
forming a resonant cavity with a total length of 30~cm. 
\label{Examples}}
\end{figure}   

The tunneling time raised at the front of opaque barriers is constant and independent of barrier length. Thus 
with increasing barrier length the signal velocity increases 
at the same rate as the length. This phenomenon is often called Hartman effect~\cite{Olkhovsky,Enders2}.

\begin{figure}[hbt]
\begin{minipage}[l]{0.50\linewidth}
\includegraphics[width=0.85\linewidth,clip=]{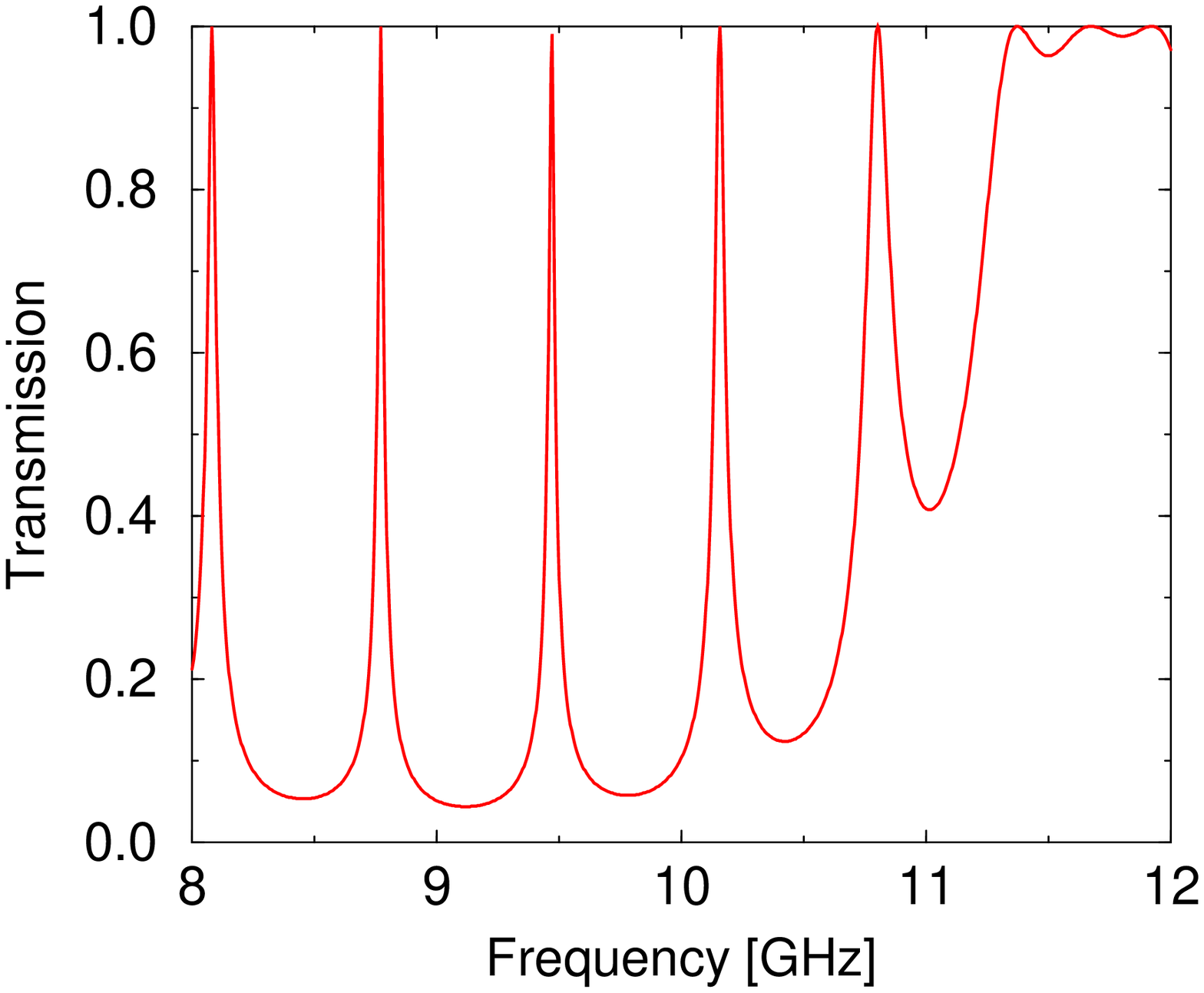}
\begin{center}
      (a)
\end{center}
\end{minipage}
\begin{minipage}[r]{0.5\linewidth}
\includegraphics[width=0.85\linewidth,clip=]{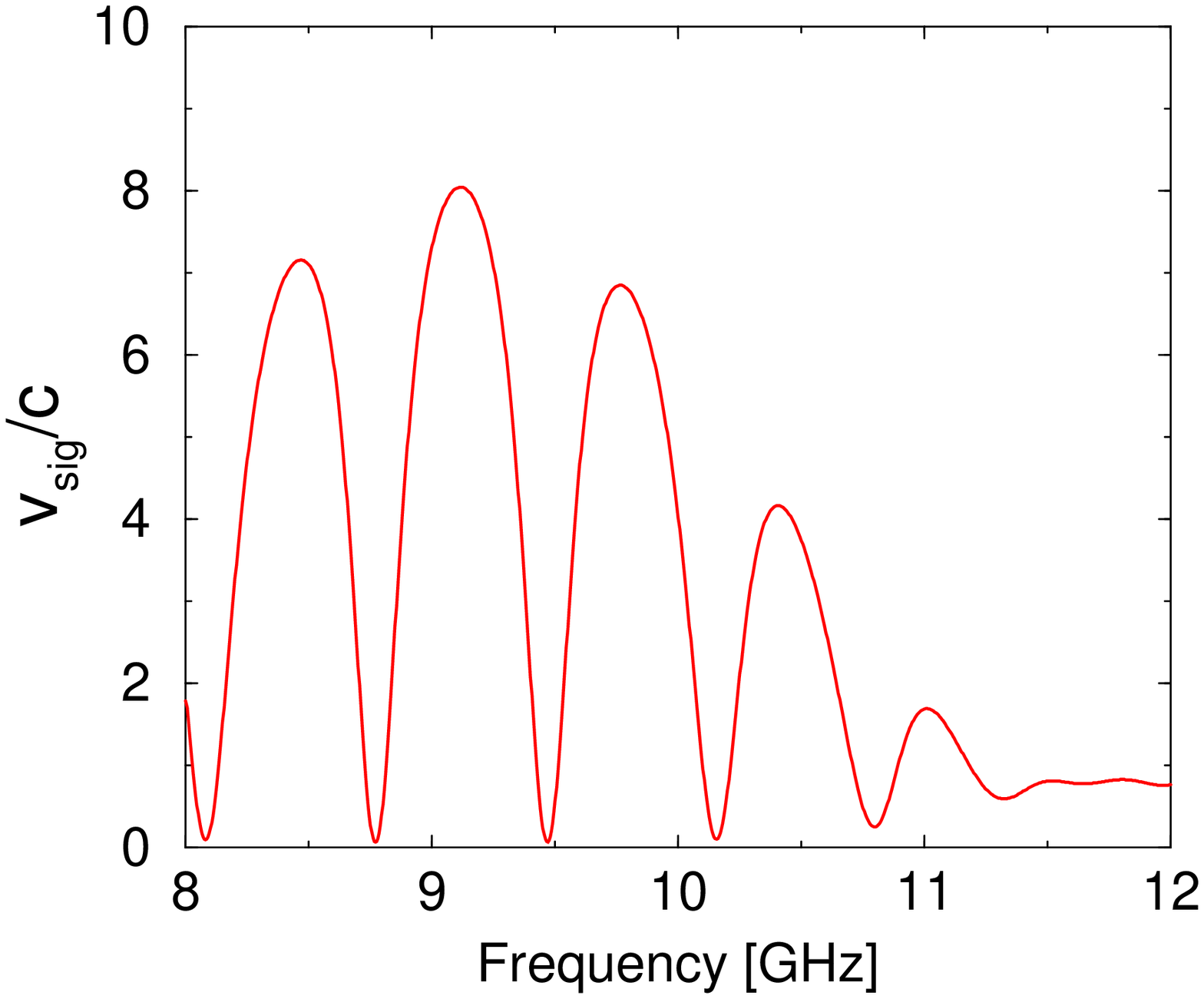}
\begin{center}
      (b)
\end{center}
\end{minipage}
\caption[]{The graph (a) shows the dispersion relation for the
resonant heterostructure vs frequency (Fig.\ref{Examples}(a)).
The transmission dispersion of the periodic heterostructure displays
forbidden gaps separated by resonant peaks. The forbidden frequency gaps 
correspond to the 
tunneling regime, for details see Ref.~\cite{Nimtz1}. The evanescent
regime is characterized by a strong attenuation due to reflection. In (b) the group (signal) velocity in units of $c$ is displayed for the
resonant periodic dielectric quarter wavelength heterostructure vs 
frequency.
\label{trans-sig}}
\end{figure}

The effective barrier length can be significantly 
lengthened by resonant barrier 
structures without decreasing the transmission. Resonant tunneling structures with forbidden frequency bands are  advantageous to 
speed-up signals with a narrow frequency band width~\cite{Nimtz1,Longhi2}. 
Figure~\ref{Examples} displays a resonant 
barrier built of two photonic lattices barriers. 
The dispersion relations of the respective transmission coefficients
and the signal velocity of the periodic dielectric quarter wavelength
heterostructure are displayed in Fig.\ref{trans-sig}. For narrow frequency band limited signals there is no 
significant dispersion effect if the carrier frequency is placed in the centre of a forbidden frequency gap. 

\begin{figure}[hbt]
\vspace*{-0.5cm}
\begin{center}
\includegraphics[angle=0,width=0.8\linewidth]{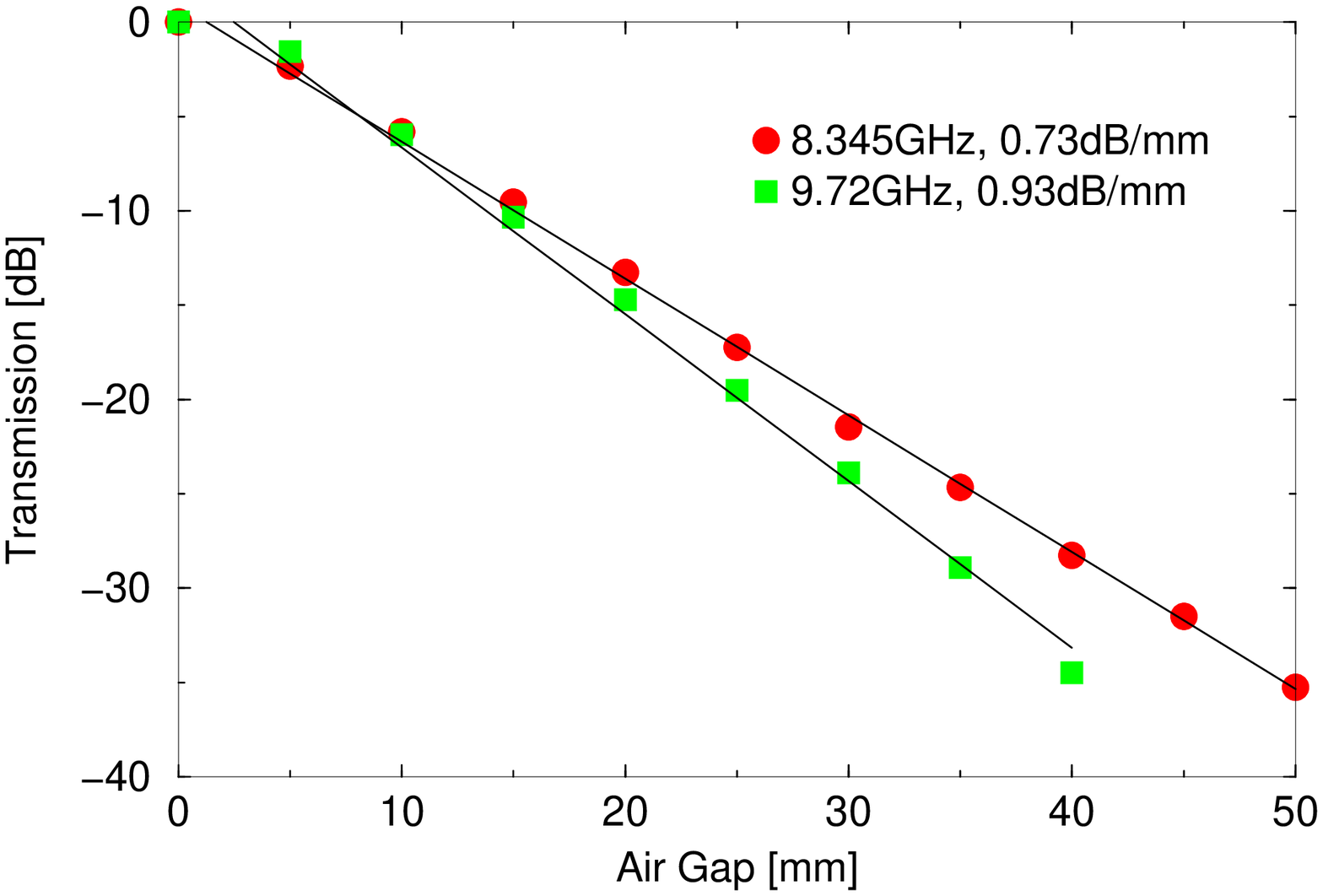}
\vspace*{-0.4cm}
\caption{Transmission vs air gap measured at two frequencies \cite{Haibel2}. 
The data follow the theoretical relation of Eqs.\ref{E},\ref{kappa}.
\label{transmission} }
\end{center}
\end{figure}

Each of the three barriers introduced in Fig.\ref{Barrier} has a different dispersion relation. A simple 
one describes the frustrated total internal reflection of a double prism. In this case the transmitted electric field $E_{t}$ and the imaginary
wave number $\kappa$ are given by the relations~\cite{Feynman}: 
\begin{eqnarray}
E_t & = &  E_0 e^{(i\omega t -\kappa z)} \label{E}\\
\kappa & = & [\frac{\omega^2}{c^2} ((\frac{n_1}{n_2})^2 - 1) sin^2 \theta)]^{1/2}, \label{kappa}
\end{eqnarray}

where $\theta$ is the angle of the incident beam, $E_{0}$ the electric field at the barrier entrance, $n_1$ and $n_2$ are the refractive indices,
and ($n_1$/$n_2$ sin$\theta $ $>$   1. The transmission $t = E_{t}/E_{0}$ as a function of air gap of a double prism was measured 
with microwaves  and is shown in Fig.\ref{transmission}. The displayed data are in agreement with Eqs.3, 4.

The tunneling time in the case of frustrated total internal 
reflection (FTIR) has been revisited  recently \cite{Stahlhofen,Haibel2}. There is a 
theoretical shortcoming in describing the time behaviour of FTIR which  
is based on the approach with ideal plane waves. This approach 
holds for an unlimited beam diameter, but is not mimicking properly 
the experimental 
procedure with limited beam diameters \cite{Haibel2,Hupert}. 

\begin{figure}[hbt]
\vspace*{-0.5cm}
\begin{center}
\includegraphics[angle=-90,width=0.8\linewidth]{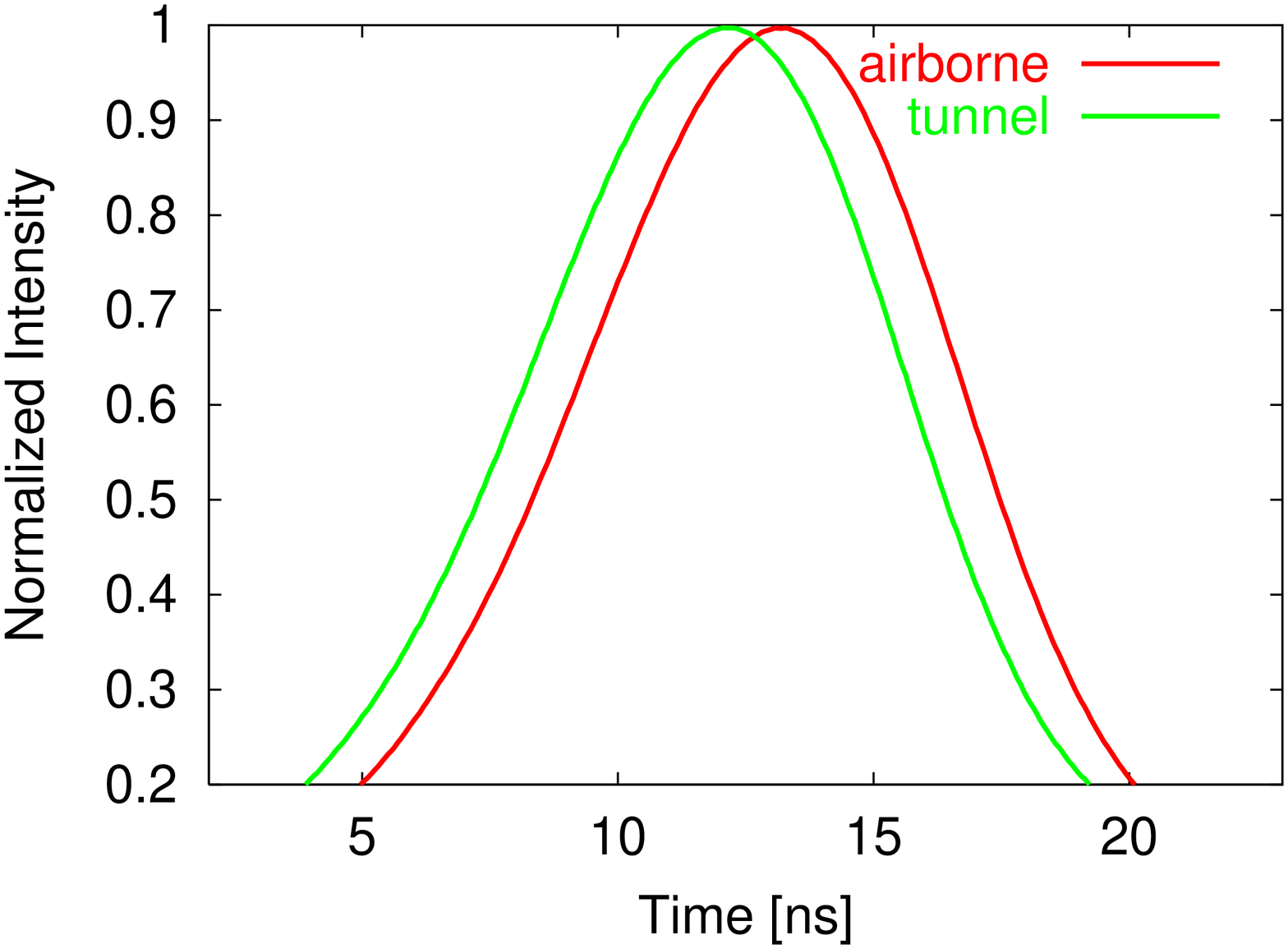}
\vspace*{0.4cm}
\caption{Measured propagation time of two signals.
The faster one has tunneled in the forbidden gap of the photonic barrier
of the length of 28~cm. The pulse´s magnitudes are normalized. The
tunneled signal (the half--width of the pulse, representing one bit) traversed the barrier
more than 800~ps faster than the airborne signal. The corresponding velocity
of the tunneled signal was 8$\cdot$c .
\label{impuls} }
\end{center}
\end{figure}

\begin{figure}[hbt]
\vspace*{-0.5cm}
\begin{center}
\includegraphics[width=1\linewidth]{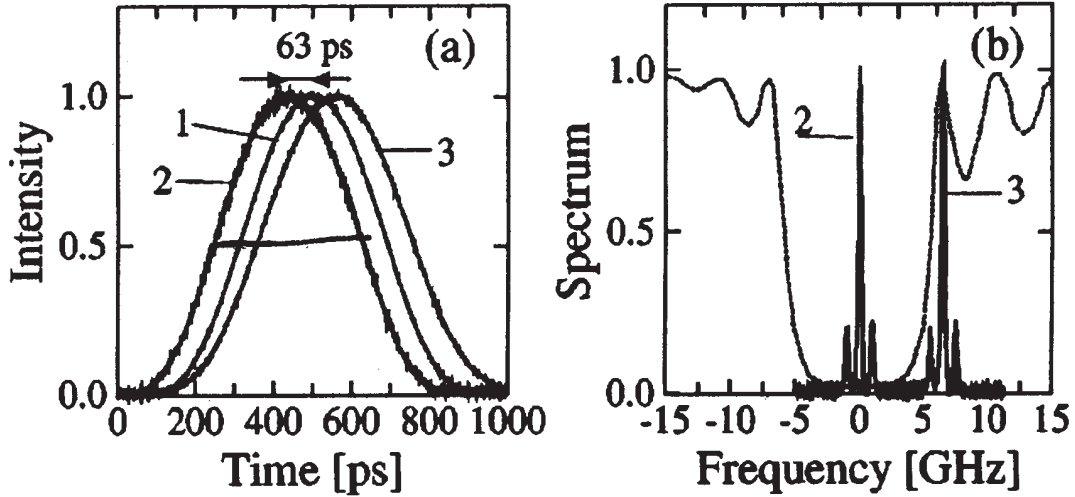}
\vspace*{-0.4cm}
\caption{Measured propagation time of three digital signals \cite{Longhi}. (a) Pulse trace 1 was recorded in vacuum. 
Pulse 2 traversed a photonic lattice in the center of the frequency band gap (see part 
(b) of the figure), and pulse 3 was recorded for the pulse 
travelling through the fiber outside the forbidden band gap. The photonic lattice was a periodic dielectric hetero-structure 
fiber.
\label{longhi} }
\end{center}
\end{figure}

In this report  two simple experiments are introduced, which demonstrate 
superluminal signal velocity in photonic barriers. The signal
is represented by a pulse as used in digital communication
systems~\cite{Nimtz4,Nimtz3}; one example was measured with microwaves and one in the infrared frequency range as 
shown in Figs. \ref{impuls}, \ref{longhi}.
 
The digital signal displayed in Fig. \ref{impuls}  tunneled with a frequency band width 
of less than $10^{-2}$  
at a speed of $8~c$.   
The carrier frequency was  9.15~GHz. 
The starting of detecting 
the  envelope of the tunneled  
signal is about 0.8 ns ealier as that of the airborne puls. The signal frequency band contained essentially evanescent components only.

\begin{figure}[hbt]
\begin{center}
\includegraphics[width=0.9
\linewidth,clip=]{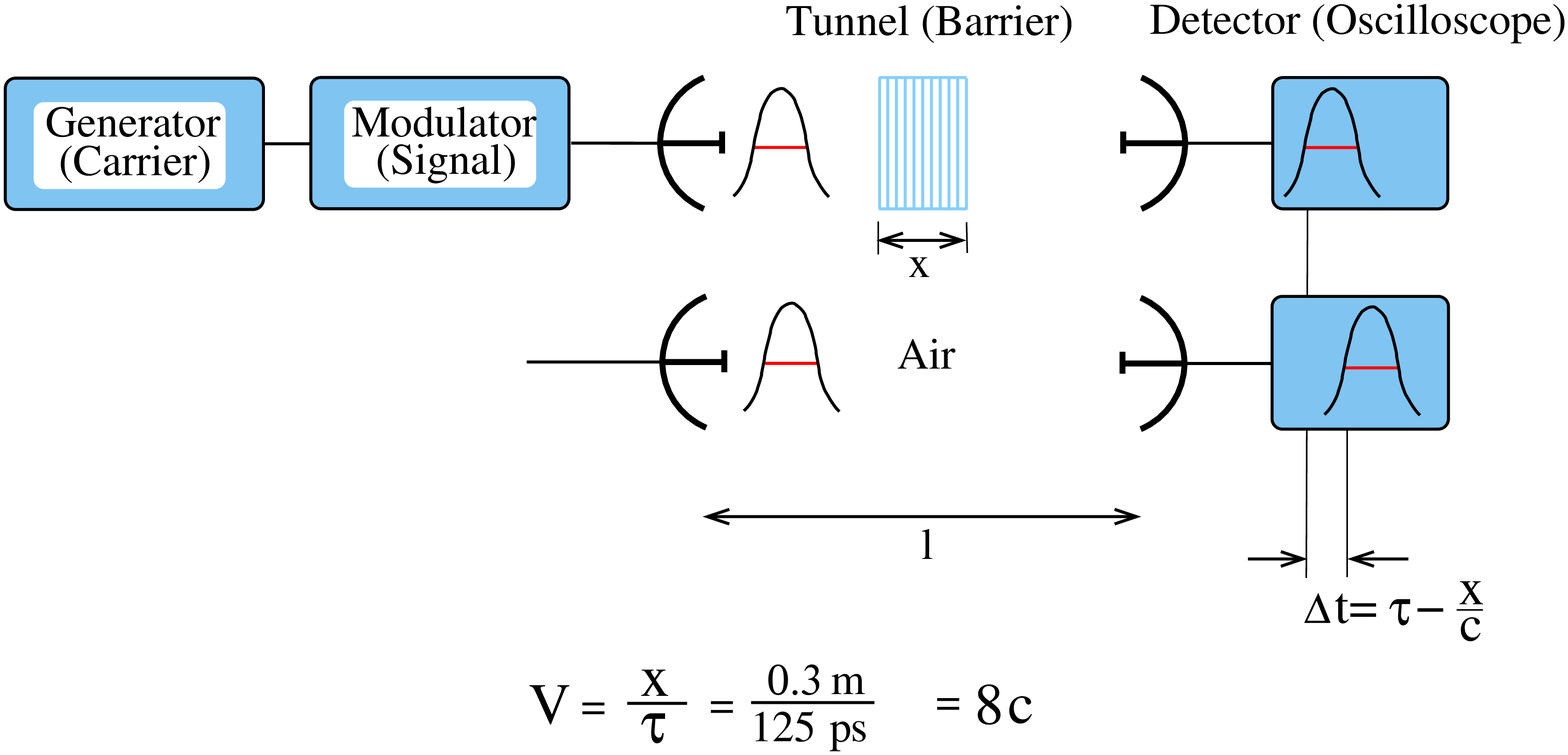}
\caption{ Experimental set-up for the periodic dielectric quarter
wavelength heterostructure to measure the group velocity, i.e. the signal velocity. \label{aufbau}}
\end{center}
\end{figure}
%

\begin{figure}[hbt]
\begin{center}
\includegraphics[width=0.8\linewidth,clip=]{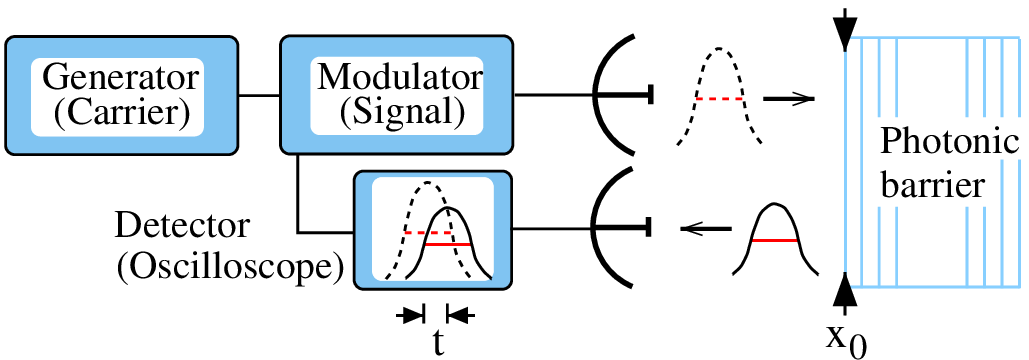}
\caption{ Experimental set--up for the periodic dielectric quarter
wavelength heterostructure to measure partial reflection as a function of barrier length.  
\label{aufbau2}}
\end{center}
\end{figure}

The experimental set-up to determine and to demonstrate superluminal group velocity 
is shown in Fig.\ref{aufbau} and \ref{aufbau2}. The group arrival was 
measured in vacuum where group, energy, and signal velocities are equal to c \cite{Brillouin,Stratton}, (actually, a signal 
does not depend on its magnitude as illustrated in Fig.\ref{Paket}).  
Amplitude modulated microwaves with a frequency of 9.15 GHz
($\lambda$ = 3.28~cm) are generated with an HP 8341B synthesized sweeper (10 MHz - 20 GHz). A parabolic
antenna transmitted  parallel beams. The transmitted signal has been received
by another parabolic antenna, rectified by a diode (HP 8472A (NEG)) and
displayed on an oscilloscope (HP 54825A).   

The propagation time of a signal was measured across the air distance
between transmitter and receiver  and across the same distance but partially filled with the
barrier of 28~cm length. The barrier structure is formed by quarter wavelength slabs
of perspex and is introduced and analyzed in Figs.~\ref{Examples}, \ref{trans-sig}.  
Each slab is 0.5~cm thick and the distance between two slabs is 0.85~cm.
Two structures are separated by an air distance of 18.9~cm
forming a resonant tunneling structure~\cite{Nimtz1}.
Comparing the two travelling  times we see that the tunneled signal arrived the detector about 
900~ps earlier than that pulse which travelled the same distance through air. 
The result corresponds to a signal velocity of the tunneled pulse of 8$\cdot$c. 

The performed measurements are asymptotic. 
There is no coupling between the generation process, the
detection process, and the photonic barrier. In addition the experiment 
is not stationary and the signal is measured in the dispersion free vacuum. 
The experimental situation is the same as that performed in the
Hong-Ou-Mandel interferometer, in which the measurement is
also asymptotic and yields the group velocity, the energy and the signal velocity 
at the same time~\cite{Nimtz5,hong}. 
An infrared example of superluminal signal velocity is displayed in Fig.\ref{longhi}.  

\begin{figure}[hbt]
\begin{center}
\includegraphics[width=0.2\linewidth,clip=]{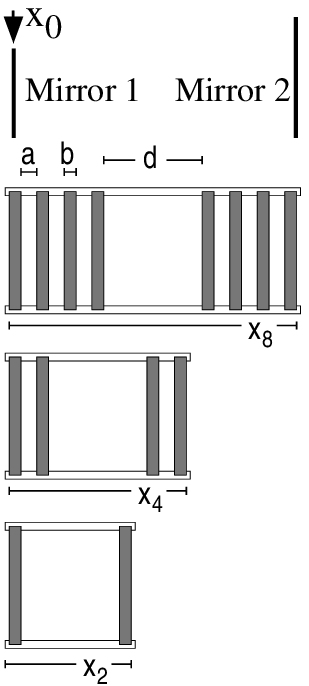}
\caption{ Experimental procedure to measure partial signal reflection depending on photonic lattice structure.
\label{aufbau3}}
\end{center}
\end{figure}

So far we have discussed transmission experiments only. 
An experimental set-up for measuring the partial reflection by photonic barriers at microwave frequencies  
is presented in Fig.\ref{aufbau2}.  
The procedure of varrying the barrier length in such an experiment is sketched in Fig.\ref{aufbau3}.

\section{Signals}
\begin{figure}[htbp]
\begin{center}
      \includegraphics[width=0.6\textwidth]{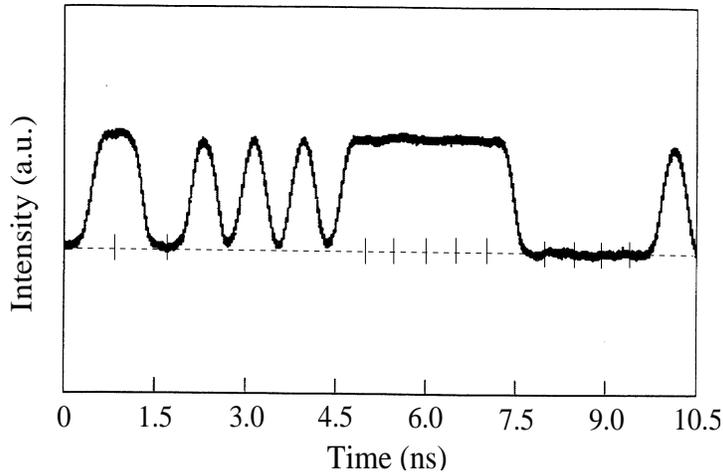}
\end{center}
      \caption{Signals:  Measured signal in arbitrary units. 
                       The half width in units of 0.2 ns corresponds to the number of bits. From left to 
                        right: 1,1,0,0,1,0,1,0,1,0,1,1,1,1,1,1,1,.....
                        The infrared 
                       carrier frequency of the signal is 
$2 \cdot 10^{14}$~Hz (wavelength 1.5~$\mu$m). The frequency-band-width of the signal is about 
$2\cdot 10^{10}$~Hz corresponding to a relative frequency-band-width of  $10^{-4}$
 \cite{Desurvire}. 
 \label{bit}}
\end{figure}

For instance, a signal may be a photon, which excites an atom with a well defined energy and polarization 
or it may be a word, which informs the receiver. 
Both examples are described by wavepackets of limited frequency band width and of limited time duration. 
The envelope of the photon as well as that of the word are travelling at the speed of light in vacuum. 
In vacuum group, signal, and energy velocities are equal c. 
Physical and of course all technical signals like those displayed in Figs.~\ref{impuls},\ref{longhi},\ref{bit} 
are frequency band limited. Technical signals are either amplitude modulated (AM)  
or frequency modulated (FM).  Definitions of the frequency bandwidth, of the time duration, and of 
the bandwidth-time 
interval product are introduced and explained in Ref.~\cite{Berkeley} for example. 
Frequency components of a signal outside the band width in charge of a hypothetical signal front 
are usually smaller than -60 dB compared with the signal peak component.

The Fourier transform yields for frequency band limited signals an unlimited time extension and hence 
a noncausal behaviour. 
However, the noncausal time components 
have never been detected. In the case of an unlimited frequency band the wave packet 
is presented by an analytic function and 
the information contained in the forward tail of the packet determines the whole packet \cite{Low2}.  
This is mathematically correct but not  relevant for signals from the physical point of view .  

As mentioned above engineers have transmitted frequency band and time duration limited signals 
with the multiplex technology already a hundred years ago. A historic picture of such a multiplex 
transmission system is shown in Fig.\ref{telegraf}. Obviously, the five  signals transmitted over 
one guide are frequency band limited 
and time duration limited. In this example the frequency band width has been 2 kHz and 
the time length about 0.3 ms. 
Shannon's and many other's theoretical investigations yielded in the result that the product 
of frequency band and of time duration represents the amount of information. The Fourier transform of 
such a multiplex technique 
yields a noncausal behaviour. This indicates that noncausal time components obtained from 
Fourier transform are not detectable~\cite{Nimtz4,Nimtz5}.

\section{Universal Tunneling Time} 
An analysis of different experimental tunneling time data obtained with opaque barriers (i.e. $\kappa~z$ $>$ 1) pointed to a universal time \cite{Haibel1}. 
The relation
\begin{eqnarray}
 \tau \approx 1/\nu = T\\
\tau \approx h/W,
\end{eqnarray}

 was found independent of frequency and of the type of barrier studied \cite{Haibel1,Esposito}, 
 where $\tau$ , $\nu$ and $T$ are the tunneling time, the carrier 
frequency or a wave packet's energy W divided by the Planck constant h, and T the oscillation time of the wave.  
The microwave experiments near 10~GHz displayed a tunneling time of about 100~ps, experiments in the 
optical frequency regime  near 427~THz a tunneling time of 2.2~fs for instance. 
In Ref.~\cite{Haibel1} it was conjectured that the relation holds also for wave packets with a rest mass having 
in mind the mathematical analogy between the Helmholtz and the Schr\"odinger equations. 
Quantum mechanical studies point to this conjecture \cite{Hartman,Collins,Low}. Recently electron tunneling time 
was measured in a field emission experiment \cite{Sekatskii}. The measured times are between 6 fs and 8 fs. Assuming an 
electron energy of 0.6 eV  (the barrier height was 1.7 eV) 
the empirical Eq.(5) yields a tunneling time of 7 fs.

\section{Photonic Applications}
Tunneling transmission has an exponential decrease with barrier length,    
the transmission  loss is due to reflection. Actually, the transmission loss is not converted into heat and may be 
recycled in a special circuit design.  

\subsection{Tunneling}
a) Recently 
Longhi et al.~\cite{Longhi} performed  tunneling of narrow band infrared pulses 
over 
a distance of  20 000 wavelenghts corresponding to about 80 000 quarter 
wavelengths layers. Results are presented in Fig.\ref{longhi}.  The overall distance of the photonic fiber lattice was 2 cm. 
(Scaling the barrier length to 10~GHz microwaves the 
barrier would  be 400~m long.) The periodic 
variation of the refractive index along the fiber between the two different quarter wavelength layers is only 
of the order of $10^{-4}$.  The measured group velocity was 2 c and the transmission intensity 
of the barrier was 1.5 \%.  

\vspace{1cm}

b) An analogous tunneling barrier of 16.81 m length is under construction at the University of Koblenz. The 
long structure is designed for microwave signals at a frequency of  9.15 GHz, i.e. at a wavelength of 3.28 cm. 
159 dielectric layers differing in the refractive index between 1.00 (air) and 1.05 of a plastic material. 
The transition time of the huge barrier will be 14 ns compared with a 
vacuum time of  56 ns. The expected signal velocity will be 4~c at a transmission of intensity of 0.16~\%~\cite{Vetter2}. 
The tunneled signal will arrive at the detector 49 ns earlier than the airborn one.

\subsection{Partial Reflection By Photonic Barriers}

a) Photonic barrier reflection 
is used  
at 1.5 $\mu m$ wavelength in fiber optics. Barriers are performed by a 2 cm long piece of glass fiber with a weakly periodically 
changed refractive index similar to the barrier used in the superluminal transmission experiment by 
Longhi et al. mentioned above \cite{Longhi}. 
The losses of reflection by a photonic barrier (imaginary impedance) 
are less than that of a metal. Photonic barriers represent more effective mirrors than metallic ones. For example photonic barriers are profitably 
used to stabilize infrared laser diodes in optoelectronics.  
\vspace{1cm}

\begin{figure}[htbp]
\begin{center}
      \includegraphics[angle=180,width=0.7\textwidth]{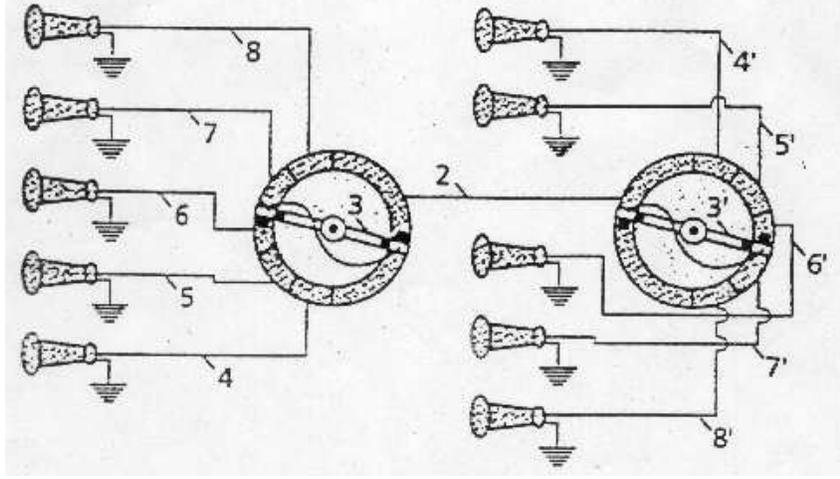}
\end{center}
      \caption{Historical picture of a multiplex transmission system Ref.~\cite{Telegraf}. 
 \label{telegraf}}
\end{figure}
\begin{figure}[hbt]
\begin{center}
\includegraphics[width=0.7\linewidth,clip=]{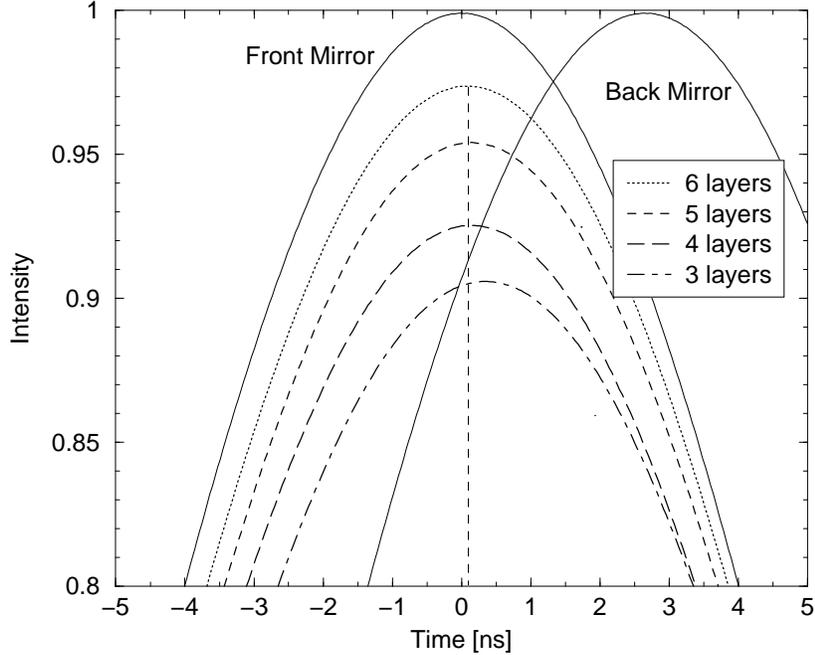}
\caption{Measured partial reflected microwave pulses vs time. Parameter is the barrier composition as shown in Fig.~\ref{aufbau3}. 
The signal reflections from metal mirrors either substituting the barrier's front or back position are displayed \cite{Vetter}.  In this 
example the wavelength has been 3.28 cm and the effective barrier length was 41 cm.
\label{nonlocal}}
\end{center}
\end{figure}


b) Figure~\ref{nonlocal} shows time dependent reflection data by two mirrors at different positions and by photonic barriers of different lengths. 
Only the magnitude of the reflected pulse is changed but not the 
reflection time by the barrier length. The observed reflection time of about 100 ps equals the tunneling time in transmission of the barrier. 
The nonlocal behavior of tunneling modes gives 
 the
 information on barrier length within one oscillation time 
 at the barrier front.

We have designed an ultrafast modulator on the basis of partial reflection. 
The effective barrier length is modulated by an electric field induced change of the refractive index 
at half of the total barrier length. This results in an amplitude change of reflection, see Fig.~\ref{nonlocal}. 
Another type of modulation can be achieved due to a local change of refractive index 
by signals exciting an optical active dielectric medium. 
For example this principle 
has been applied in experiments on negative group velocity, see e.g. Ref. \cite{Wang}. 
In the case of the above microwave experiment the modulations at the distance of 15 cm away 
from the barrier entrance appears at the barrier front within 100~ps, whereas the corresponding 
luminal propagation time is five times longer.

\section{Electronic Applications}

\subsection{The Tunneling Diode}
The first man made tunneling device was the tunneling diode. It was invented by Esaki around 1960. 
This nonlinear electronic device is more and more used since that time.
However, the tunneling time
which would give the ultimate dynamical specification of such a diode  
has never been measured yet. Our conjecture is: the universal photonic tunneling time Ref.~\cite{Haibel1} is valid   
also for the electronic tunneling process. Actually, recent electronic tunneling time experiments support the conjecture~\cite{Sekatskii}. 
The experimental data is 
in agreement with relation Eq. 6 .

\vspace{1cm}
 
\subsection{Superluminal Electron 
Transport}
It was shown in several quantum mechanical studies by Low and Mende for instance that  {\it a particle suitably localized in space and time, which is transmitted 
through a long, high barrier, travels as if it tunneled it in zero time.}~\cite{Low}.  Of course, the 
time spent inside the barrier was considered only. Again as in the case of photonic tunneling the barrier traversal velocity was 
superluminal even in the case of relativistic approaches~\cite{Collins,Low,Leavens}.  

The 
electronic transport in a semiconductor is rather slow compared with the velocity of light. The utmost highest electron 
velocity is given by 
the ballistic electron transport like in the case of an electron microscope or some semiconductor nano-device structures.  
Bias voltages of electronic devices are of the order of 1V. This results in a ballistic electron velocity of  
the order of magnitude of $10^6$ m/s, which is two orders of magnitude smaller than c.

\subsubsection{Electronic Lattice Structures}
We propose an electronic lattice structure with alternating quarter wavelength layers of slightly different bandgaps, 
which can be traversed 
at superluminal speeds. The conduction band electron wavelength is of the order of magnitude of 1 nm. Ultrafast coupling of 
electronic device elements in a circuit could be performed and 
accelerate the speed of computers.  For instance a periodic structure of Si/SiGe quarter wavelength layers 
represents an electronic lattice. 
Such a doping  of a Si-semiconductor structure 
with the SiGe alloy yields a weak variation of the band gap 
analogous to the periodic dielectric fiber structure mentioned above. 
The electronic structure could have extensions up to more than 1~$\mu$m and could be used to perform 
ultra-fast interconnections between device elements. 

\subsubsection{pn-Tunnel Junctions}
Interband tunneling the basis of the classical tunneling diode can also be used for fast electronic interconnections. By an 
appropriate doping profile the tunneling path can be adjusted between some 100~nm up to several 1000~nm. 

There is a problem left with all the tunneling applications: the high reflection at the barrier entrance. However, 
tunneling is not a dissipative process with energy loss. The reflected electronic power should be recycled by a smart circuit design.

\section{Summing up}

The tunneling process shows amazing properties in the case of opaque barriers  
to which we are not used to from classical physics.
The tunneling time is universal and arises at the barrier front. 
It 
equals approximately the reciprocal frequency of the carrier frequency or of the wave packet energy divided by the Planck constant h.   
Inside a barrier the wave packet does not spent any time. This property results in superluminal signal and energy velocities.  The latter  
became obvious in the single photon experiment, where the detector measured the superluminal energy velocity of the photon \cite{Steinberg2}. 
Another strange experience is that evanescent fields
are not fully describable by the Maxwell equations. They carry a ${\it negative}$ 
energy for instance which makes it impossible to detect them~\cite{Bryngdahl,Nimtz3,Gas} and they are nonlocal. Incidentally, 
the properties are in agreement with the wave mechanical
tunneling. For instance, this is a situation similar to the Hydrogen atom and the photoelectric effect, 
where quantum mechanics is necessary to explain the atom's stability and the photon-electron interaction.   

The energy of signals is always finite resulting in a limited frequency 
spectrum. This is a consequence of Planck's quantization of radiation with
an energy minimum of $\hbar \omega$~\cite{Nimtz4}. An electric field cannot be measured directly. 
All detectors 
need at least one energy quantum $\hbar \omega$ in order to respond. This 
is a fundamental deficiency of classical physics, which assumes 
any small amount of field and charge is measurable. 

A physical signal
has not a well defined envelope front. The latter would need infinite high frequency components
with a accordingly high energy~\cite{Papoulis,Nimtz5}. In addition signals are not presented by an analytical function, 
otherwise the complete information would be contained in the forward tail of the signal~\cite{Low2}. 

Another consequence of the frequency band limitation of signals is, if
they have only evanescent mode components, they can violate Einstein
causality, which claims that signal and energy velocity is $\le$ c.   

In spite of so much arguing about violation of Einstein causality~\cite{Chiao}, 
all the properties introduced above are useful for novel fast devices, for both photonics and electronics.

According to Collins et al.~\cite{Collins}  the disputes on zero tunneling time (the time spent inside a barrier) 
are redundant after reading the papers by Wigner and Hartman.  
The discussions about superluminal tunneling remind me to the problem of multiplex transmission displayed in Fig.~\ref{telegraf}. Here 
a signal's finite time duration and  
frequency band limitation violate causality according to Fourier transform. However, no one had a ringing-up 
before the other phone was switched on. This indicates the crucial role of finite frequency bands and finite time 
duration of signals without violating the principle of causality \cite{Nimtz5}.

\section{Acknowledgments}
I gratefully acknowledge discussions with A. Haibel, A. Stahlhofen, and R.-M. Vetter.

\end{document}